\documentclass[letterpaper, 10 pt, conference]{ieeeconf}  

\IEEEoverridecommandlockouts                              

\usepackage{graphicx}
\usepackage{url}  
\usepackage{pifont}

\usepackage{amsmath}

\usepackage{lipsum}

\title{\LARGE \bf
DIN-CTS: Low-Complexity Depthwise-Inception Neural Network with Contrastive Training Strategy for Deepfake Speech Detection  
}

\author{Lam~Pham$^{*}$, 
        Dat~Tran$^{*}$,
        Phat~Lam, 
        Florian~Skopik, \\
        Alexander~Schindler, 
        Silvia~Poletti,
        David~Fischinger,
        Martin~Boyer       
\thanks{L. Pham, F. Skopik, A. Schindler, S. Poletti, D. Fischinger, and M. Boyer are with Austrian Institute of Technology, Vienna, Austria} 
\thanks{D. Tran is with FPT University, Vietnam} 
\thanks{P. Lam is with HCM University of Technology, Vietnam} 
\thanks{(*) Main and equal contribution into the paper.}
}

\begin{document}

\maketitle
\thispagestyle{empty}
\pagestyle{empty}

\begin{abstract}
In this paper, we propose a deep neural network approach for deepfake speech detection (DSD) based on a low-complexity Depthwise-Inception Network (DIN) trained with a contrastive training strategy (CTS) (This work is a part of our DERAME FAKES and EUCINF projects). 
In this framework, input audio recordings are first transformed into spectrograms using Short-Time Fourier Transform (STFT) and Linear Filter (LF), which are then used to train the DIN. 
Once trained, the DIN processes bonafide utterances to extract audio embeddings, which are used to construct a Gaussian distribution representing genuine speech. 
Deepfake detection is then performed by computing the distance between a test utterance and this distribution to determine whether the utterance is fake or bonafide.
To evaluate our proposed systems, we conducted extensive experiments on the benchmark dataset of ASVspoof 2019 LA.
The experimental results demonstrate the effectiveness of combining the Depthwise-Inception Network with the contrastive learning strategy in distinguishing between fake and bonafide utterances. We achieved Equal Error Rate (EER), Accuracy (Acc.), F1, AUC scores of 4.6\%, 95.4\%, 97.3\%, and 98.9\% respectively using a single, low-complexity DIN with just 1.77 M parameters and 985 M FLOPS on short audio segments (4 seconds). Furthermore, our proposed system outperforms the single-system submissions in the ASVspoof 2019 LA challenge, showcasing its potential for real-time applications.

\indent \textit{Items}--- deepfake audio, spectrogram, feature extraction, classification model.
\end{abstract}
\section{INTRODUCTION}
\label{intro}
Thanks to advancements in deep learning technologies, speech
generation systems now beneficially support a wide range of real-world applications including text-to-speech for individuals with speech disorders, voice chatbots in call centers, cross-linguistic speech translation, etc.
However, deep learning has also facilitated the creation of fake speech for malicious purposes, such as spoofing attacks, raising serious security concerns. As a result, deepfake speech detection (DSD) has recently gained significant attention from the research community.
%
The state-of-the-art systems, which have been proposed for the task of deepfake speech detection, approach  neural network architectures and deep learning techniques~\cite{survey_01, survey_02, survey_03}.
To achieve a high-performance DSD system, ensembles of input features or models are leveraged.
In particular, multiple input features of LFCC, PSCC, LLFB were used in~\cite{m06}.
Similarly, authors in~\cite{lam_01} made uses of MEL, CQT, GAM, LFCC,  and Wavelet based spectrograms.
Regarding model ensembling, two approaches are commonly used. The first involves fusing the individual results of multiple models, as seen in~\cite{m09} with the use of LCNN and ResNet. 
The second approach integrates multiple branches within a single network architecture, where feature maps extracted from different pre-trained models are combined. For an example, feature maps from XLS-R, WavLM, and Hubert are merged in~\cite{m15}.
%
Although ensemble models prove effective to achieve high performance (i.e., the best systems proposed in deepfake speech detection challenges of ASVspoof 2019~\cite{asv19}, 2021~\cite{asv21}, and 2024~\cite{asv24} leveraged ensemble models), this method leads an issue of a high-complexity model with a large number of trainable parameters and FLOPS.
This poses a challenge for applying ensemble models to real-world applications that require a real-time inference or are constrained by hardware limitations.
\begin{figure*}[t]
    \centering
    \includegraphics[width =1.0\linewidth]{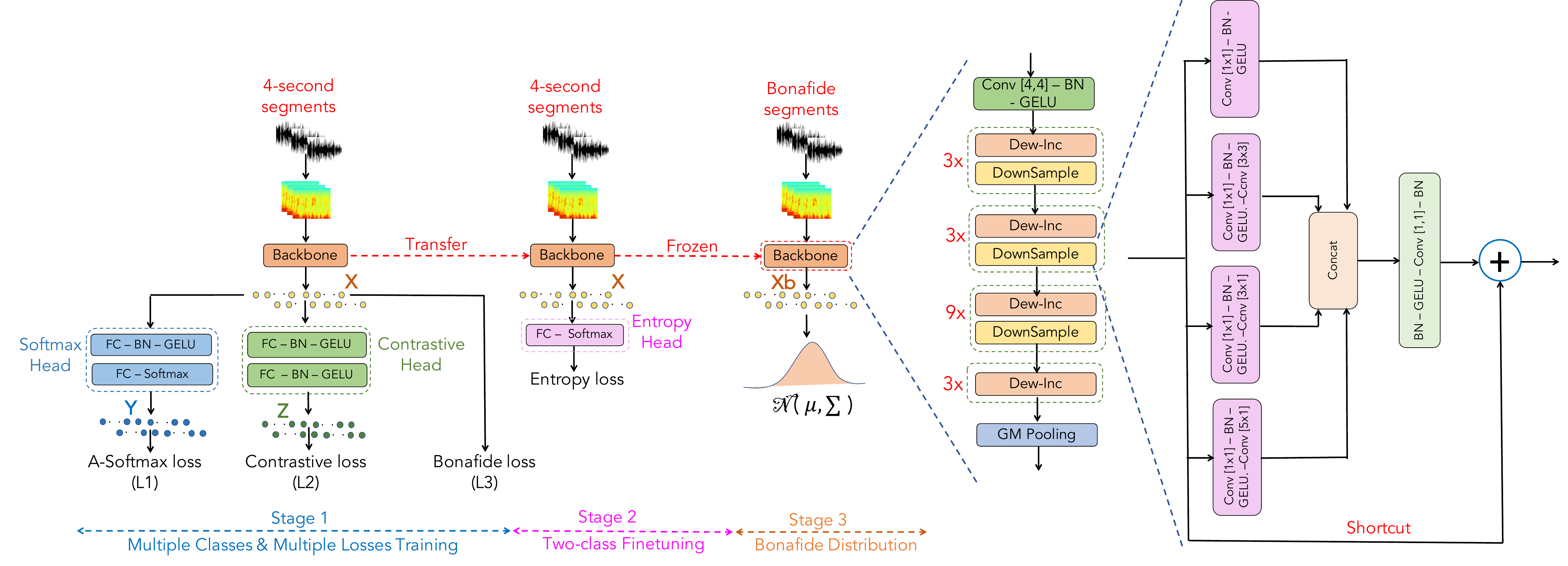}
           	\vspace{-0.2cm}
	\caption{The proposed Depthwise-Inception Network architecture and Contrastive Training Strategy}
    \label{F1}
\end{figure*} 
Recently, single models with encoder-decoder based architectures have become popular~\cite{survey_01}.
In these systems, encoder architectures leverage pre-trained models such as Whisper~\cite{whisper}, WavLM~\cite{wavlm}, or Wave2Vec2.0~\cite{wav2vec20} (i.e., these models were trained on large-scale datasets of human speech in advance) to extract general feature maps.  
Meanwhile, decoders present diverse architectures such as Multilayer Perceptron~\cite{lam_01}, GAN-based architecture~\cite{m19}, Multi-feature attention~\cite{m20}, Graph Attention Network~\cite{m34} to explore the feature maps extracted from the encoders.
Although this approach proposes single models for the DSD task, leveraging pre-trained models as encoders still results in a highly complex system. For examples, the smallest Whisper model presents 39 M parameters and the largest Whisper model has 1550 M parameters. Meanwhile, the pre-trained Wave2Vec2.0 BASE and LARGE models presents around 95 M and 300 M of parameters, respectively.
Additionally, this approach mainly focuses on exploring encoder and decoder architectures rather than analyzing training strategies which enforce the model to separate the distributions of bonafide and fake utterances.

To tackle these mentioned limitations, we propose a deep-learning-based model for the deepfake speech detection (DSD) and highlight the contributions: 
    (1) We first propose a novel and low-complexity deep neural network for DSD task that is inspired by depthwise convolution and inception architectures, referred to as the Depthwise-Inception Network (DIN).
    (2) To train DIN model, we propose a contrastive training strategy (CTS) that proves effective to separate distribution of bonafide and fake utterances. 
    (3) By combining DIN model and the CTS method, we achieved a low-complexity and high-performance DSD system.

\section{Depthwise-Inception Network Architecture and Contrastive Training Strategy}
\label{systems}


The detailed architecture of the proposed deep-learning-based model is denoted in Fig.~\ref{F1} and comprises two main parts: spectrogram feature extraction and deep learning model.

\subsection{Spectrogram Feature Extraction}
\label{spec}
The input utterances are first split into 4-second audio segments. 
This segment length generally provides sufficient context to capture important features and allows faster training and inference for applications that require real-time detection. 
Then, Short Time Fourier Transform (STFT) and Linear Filter (LF) are applied on 4-second segments to generate STFT-LF spectrograms. 
By setting window size, hop size, and linear filter number of 1024, 512, and 64 respectively, we obtain STFT-LF spectrograms of 128$\times$128.
Then, each spectrogram is concatenated with its first and second deviations to create a 3-channel representation, resulting in a size of 3$\times$128$\times$128.
Finally, we apply the SpecAug data augmentation~\cite{spec_aug} to the spectrograms before feeding into the backend classification model.

\subsection{Depthwise-Inception Network Architecture}
\label{network}
By comparing pairs of bonafide and fake utterances, we observed that the power distribution across the frequency axis in fake spectrograms follows a regular pattern, whereas in bonafide utterances, it is more variable. 
This inspired us to make use of inception network layers which are effective to learn the minor difference in local regions of the spectrograms.
Additionally, we leverage depthwise convolution layers and pointwise convolution layers rather than the traditional convolution layers to reduce the model size, but still preserve the power of convolution computation to be able to generate distinct feature maps between fake and bonafide utterances.
By combining depthwise convolution and inception layers, we construct the novel Depthwise-Inception Network (DIN).
Fig.~\ref{F1} illustrates the proposed DIN architecture, which can be divided into two main components: the backbone and the heads.
The backbone as shown in the right side of the Fig.~\ref{F1} presents a dense layer (Conv [4x4], BatchNorm (BN) - Gaussian Error Linear Units (GELU)), followed by four Dew-Inc blocks, and a final Global Max Pooling (GM Pooling) layer. 
 For each Dew-Inc block, it presents inception and residual based architecture with four branches of inception-convolution layers with different convolutional kernels (i.e., [1$\times$1], [3$\times$3], [3$\times$1], [5$\times$1]) and one residual shortcut.
%
The output of the backbone is fed into two heads: Softmax head and Contrastive head.
Softmax head presents two dense layers. 
The first dense layer presents Fully Connected layer (FC), followed by BN and GELU. 
The second dense layer comprises FC and Softmax layers.
Meanwhile, Contrastive head performs two dense layers which share the same configuration of FC, BN, and GELU.

\subsection{Contrastive Training Strategy}
\label{train}

To train the proposed DIN model, we introduce a contrastive training strategy with three training stages. 
In the first stage, we train DIN model with multiple classes and multiple loss functions.
While bonafide utterance is considered as one class, fake generators are the other classes. 
Training with multiple classes enforces DIN model to learn distinct features among bonafide and fake utterances.

To further enforce the training process, we propose three loss functions. Let $N$ be the number of spectrograms in each training batch. Then, the output feature vector of the backbone, the Softmax head and the Contrastive head are indicated as $\boldsymbol{X = \{x_1, x_2, ..., x_N\}}$, the output feature maps of Softmax head and Contrastive head as $\boldsymbol{Y = \{y_1, y_2, ..., y_N\}}$ and $\boldsymbol{Z = \{z_1, z_2, ..., z_N\}}$, respectively.
The first loss function $L_1$ is A-Softmax loss as shown in Equation~(\ref{loss_1}).
\begin{equation}
\resizebox{.9\hsize}{!}{$
L_1= \frac{1}{N} \sum_{n=1}^N -\log \left(\frac{\exp({s \phi (\theta_{c}^{n})})}
{\exp{ (s\phi (\theta_{c}^{n})}) + \sum_{j \not = c}\exp({ s\cos{\theta_j^{n}}})} \right)
$}
\label{loss_1}
\end{equation}

where $\phi(\theta) = (-1)^k \cos{(m \theta)} - 2 k$, $m$ is the margin, $s$ is the scalar factor; $\theta_{c}^{n}$ is the angle between the embedding  $\boldsymbol{y_n}$ and the class weight $\boldsymbol{w}_{y_n}$, $c$ is the label of the embedding $\boldsymbol{y_n}$. This loss is applied to the feature maps $\boldsymbol{Y}$ for detecting multiple classes of bonafide and fake generators. 
We use A-Softmax loss~\cite{Asoftmax_loss} rather than the traditional Entropy loss to maximize angles between feature maps from different classes.
$m$ and $s$ are empirically set to 4 and 30, respectively.

The second loss $L_2$ is a type of contrastive loss which is applied for self-supervised learning~\cite{Contrastive_loss}. 
To apply this loss, we consider only two fake speech generators, namely Text-To-Speech (TTS) generator and Voice Conversion (VC) generator. The loss, which is applied on the feature maps $\boldsymbol{Z}$, aims to minimize the distances among feature maps of the same classes $\boldsymbol{Z_C}$ and maximize the distances among feature maps of the different classes of $\boldsymbol{Z_C}$ and $\boldsymbol{Z_J}$, where $\boldsymbol{Z_C}$ and $\boldsymbol{Z_J}$ are subsets of $\boldsymbol{Z}$ with $C+J=N$. The loss is defined as:

\begin{equation}
\resizebox{.9\hsize}{!}{$
L_2 = \frac{1}{N}\sum_{n=1}^N \frac{-1}{C} \sum_{c=1}^C 
\log{\left(\frac{\exp({\boldsymbol z_n \cdot \boldsymbol z_c / \tau})}
{\exp({\boldsymbol z_n \cdot \boldsymbol z_c / \tau}) + \sum_{j=1}^J 
\exp({\boldsymbol z_n \cdot \boldsymbol z_j / \tau})} \right)}
$}
\label{loss_2}
\end{equation}

where $\tau$ is empirically set to 0.01. 

A key consideration is that, with the advancement of deep learning, SDS systems should be continuously updated to keep pace with emerging fake speech generators. However, given the rapid proliferation of new generators, this is often impractical. As a result, SDS models frequently encounter a large number of unseen fake speech they were not trained on, posing a challenge to their effectiveness. 
To tackle this issue, we develop a loss function which focuses on bonafide rather than fake speech.
In particular, the loss function aims to minimize the variance of the distribution of feature maps of bonafide utterances. 
The loss is presented by Equation~(\ref{loss_3})
\begin{equation}
	L_3 = \frac{1}{K} \sum_{k=1}^K ||\boldsymbol{x_k} - \boldsymbol{c}||_2^2
\label{loss_3}
\end{equation}
where $\boldsymbol{c}$ is the central feature map of bonafide utterances, $\boldsymbol{x_k}$ is a bonafide feature map obtained from the backbone, $K$ is the number of bonafide utterances in the bach of $N$ spectrograms ($K < N$). The central feature map $\boldsymbol{c}$  is the average of bonafide feature maps in each training batch, but it is recomputed from all bonafide feature maps in the entire training set every 5 epoches.

By combining three loss function, we obtain the final loss $L$ to train the DIN archtiecture as the Equation~(\ref{final_loss})
\begin{equation}
L = \alpha L_1 + \beta L_2 + \gamma L_3
\label{final_loss}
\end{equation}
where $\alpha$, $\beta$, and $\gamma$ are empirically set to 0.2, 0.4, 0.4, respectively.

\begin{table}[t]
    \centering
    \caption{Proposed Contrastive Training Strategy}
    \vspace{-0.3cm}
    \label{tab:algo}
    \scalebox{1.0}{
    \begin{tabular}{l} 
    \hline
         \textbf{Algorithm 1}: Contrastive Training Strategy \\ 
    \hline
         \textbf{\underline{Input:}} A set of $T$ spectrograms split into batches of \\
         $\boldsymbol{B_t} = \{ \boldsymbol{I}_1, \boldsymbol{I}_2, \dots, \boldsymbol{I}_N\}$ \\
         \textbf{\underline{Output:}} Trained model for fake/bonafide speech detection.\\ 
         \textbf{\underline{Components:}} \\ 
         - The backbone $E$ to extract a set of embeddings \\
         $\boldsymbol{X} = \{ \boldsymbol{x}_1, \boldsymbol{x}_2, \dots, \boldsymbol{x}_N\}$ 
         from batch $\boldsymbol{B_t}$ \\

         - The Softmax Head $SH$ to extract embeddings \\
         $\boldsymbol{Y} = \{ \boldsymbol{y}_1, \boldsymbol{y}_2, \dots, \boldsymbol{y}_N\}$ 
         from $\boldsymbol{X}$ \\

         - The Contrastive Head $CH$ to extract embeddings \\
         $\boldsymbol{Z} = \{ \boldsymbol{z}_1, \boldsymbol{z}_2, \dots, \boldsymbol{z}_N\}$ 
         from $\boldsymbol{X}$ \\

         - A set of 4 losses represented by functions: 
         $\mathcal{H} = \{ H_1, H_2, H_3, H_4\}$ \\

    \hline
    \textbf{\underline{The first stage:}}\\
    \textbf{for} $e=1$ to Training Epochs \textbf{do}:\\
    \quad \textbf{for} $t=1$ to $T/N$ \textbf{do}:\\
    \quad \quad - Extract embeddings $\boldsymbol{X, Y, Z}$ from batch $\boldsymbol{B_t}$: \\

    \quad \quad \quad $\boldsymbol{X} \leftarrow E(\boldsymbol{B_t})$ \\
    \quad \quad \quad $\boldsymbol{Y} \leftarrow SH(\boldsymbol{X})$ \\
    \quad \quad \quad $\boldsymbol{Z} \leftarrow CH(\boldsymbol{X})$ \\

    \quad \quad - Compute losses: \\
    \quad \quad \quad $L_1 \leftarrow H_1(\boldsymbol{Y})$ \\
    \quad \quad \quad $L_2 \leftarrow H_2(\boldsymbol{Z})$ \\
    \quad \quad \quad $L_3 \leftarrow H_3(\boldsymbol{X})$ \\

    \quad \quad - Compute final loss: \\
    \quad \quad \quad $L \leftarrow \alpha L_1 + \beta L_2 + \gamma L_3$ \\

    \quad \quad - Backward pass and update weights of backbone and heads \\

    \quad \textbf{end for}\\ 
    \textbf{end for}\\
    
    \hline
    \textbf{\underline{The second stage:}}\\
    \textbf{for} $e=1$ to Finetune Epochs \textbf{do}:\\
    \quad \textbf{for} $t=1$ to $T/N$ \textbf{do}:\\
    \quad \quad - Extract embeddings $\boldsymbol{X}$ from batch $\boldsymbol{B_t}$: \\
    \quad \quad \quad $\boldsymbol{X} \leftarrow E(\boldsymbol{B_t})$ \\

    \quad \quad - Compute Entropy loss: \\
    \quad \quad \quad $L_{\text{Entropy}} \leftarrow H_4(\boldsymbol{X})$ \\

    \quad \quad - Backward pass and update weights \\

    \quad \textbf{end for}\\ 
    \textbf{end for}\\
    
    \hline
    \textbf{\underline{The third stage:}}\\
    \quad - Extract bonafide embeddings $\boldsymbol{X_b}$ from all batches $\boldsymbol{B_t}$: \\
    \quad \quad $\boldsymbol{X_b} \leftarrow E(\boldsymbol{B_t})$ \\

    \quad - Compute distribution of bonafide embeddings: \\
    \quad \quad $\boldsymbol{X_b} \sim \mathcal{N}(\boldsymbol{\mu},  \boldsymbol{\Sigma})$ \\

    \hline
    \end{tabular}
    }
    \vspace{-0.4cm}    
\end{table}

\begin{table*}[t]
    \caption{Performance comparison between baseline (ResNet18) and our proposed deep learning models on ASV2019-LA} 
    \vspace{-0.2cm}
    \centering
    \scalebox{1.0}{
    \begin{tabular}{|l|c|c|c|c|c|c|} 
        \hline 
        \textbf{Network \& Settings} & \textbf{Acc}~$\uparrow$ & \textbf{F1}~$\uparrow$ & \textbf{AUC}~$\uparrow$ & \textbf{EER}~$\downarrow$ & \textbf{Parameters}~$\downarrow$ & \textbf{FLOPS}~$\downarrow$ \\  
        \hline    
        (\textbf{Baseline}) ResNet18 w/ Entropy head \& Entropy loss & 92.0\% & 95.4\% & 96.6\% & 8.0\% & 11.18M & 1192M \\

        (M1) DIN backbone w/ Entropy head \& Entropy loss & 92.1\% & 95.5\% & 97.5\% & 7.9\% & 1.77M & 985M \\

        (M2) DIN backbone w/ multiple heads, multiple losses (stage 1\&2) & 94.6\% & 97.0\% & 98.4\% & 5.3\% & 1.77M & 985M \\ 

        (M3) DIN backbone w/ multiple heads, multiple losses (3 stages) & 95.4\% & 97.3\% & 98.9\% & 4.6\% & 1.77M & 985M \\  

        \hline
    \end{tabular}
    }
    \vspace{-0.1cm}
    \label{table:R1} 
\end{table*}

\begin{figure*}[t]
    \centering
    \includegraphics[width =1.0\linewidth]{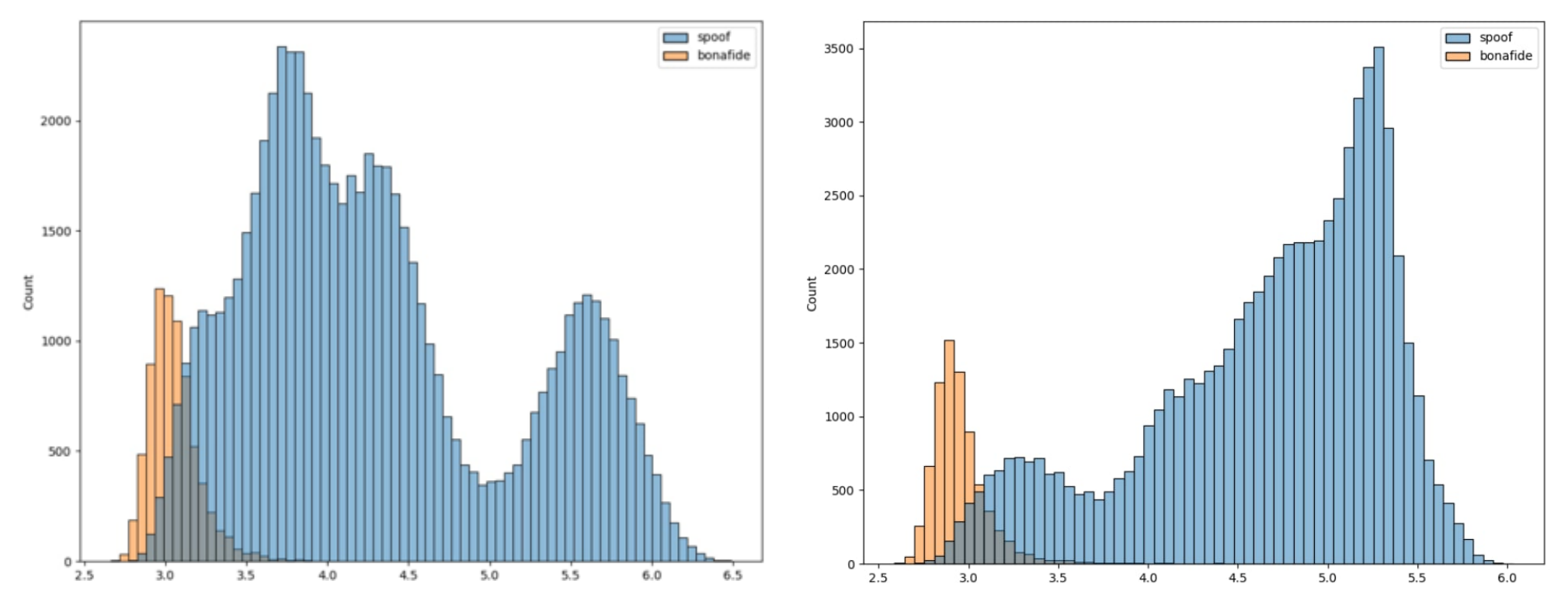}
           	\vspace{-0.2cm}
	\caption{Histogram of Mahalanobis distances between the training bonafide distribution (`Train' subset) and the test bonafide \& fake utterances (`Evaluating' subset) with the model M1 on the left side and the model M3 on the right side}
       	\vspace{-0.4cm}
    \label{F3}
\end{figure*}
In the second stage, we replace the A-Softmax and Contrastive heads by a new head, referred to as the Entropy head with only a FC layer and a Softmax layer.
While the Entropy head is fine-tuned with a high learning rate, the trainable parameters in the backbone are fine-tuned with a low learning rate at this stage.

In the third stage, we fed only bonafide utterances into the pre-trained DIN model obtained from the second stage, extracting the output feature maps of the backbone that are denoted by the $\mathbf{X_b}$ in Fig.~\ref{F1}.
Then, the mean $\boldsymbol{\mu}$ and the co-variance matrix $\boldsymbol{\Sigma}$ are computed from the feature maps $\mathbf{X_b}$.
In other words, we obtain the Gaussian distribution of bonafide utterances $\mathbf{X_b} \sim \mathcal{N}(\boldsymbol{\mu}, \boldsymbol{\Sigma})$.
In summary, the proposed contrastive training strategy is described in Table~\ref{tab:algo}.

\subsection{Inference Process}
In the inference process, an testing utterance is fed into the pre-trained DIN model in the second stage to extract the output feature map of the backbone $\boldsymbol{x^t}$. 
The Mahalanobis distance $d$ between the testing feature map $\boldsymbol{x^t}$ and the Gaussian distribution of bonafide utterances $\mathcal{N}(\boldsymbol{\mu}, \boldsymbol{\Sigma})$ is computed by Equation~(\ref{pos_equ}) to decide whether the testing utterance is fake or bonafide.
\begin{equation}
    d\{\boldsymbol{x^t}, \mathcal{N}(\boldsymbol{\mu}, \boldsymbol{\Sigma})\} = \sqrt{(\boldsymbol{x^t} - \boldsymbol{\mu})^T \boldsymbol{\Sigma}^{-1}(\boldsymbol{x^t} - \boldsymbol{\mu})}
    \label{pos_equ}
\end{equation}

\section{Experiments and Results}
\label{exper}

\subsection{Datasets and evaluation metrics}
We evaluate the proposed models on the Logic Access dataset of ASVspoof 2019 challenge (ASV2019-LA). 
%
The models are trained on the `Train' subset and evaluated on `Development' subset from ASV2019-LA. These subsets comprise bonafide utterances and fake utterances from six speech generators (2 VC and 4 TTS systems), referred to as A01 to A06.
Finally, the models are tested on the `Evaluation' subsets from ASV2019-LA. This subset comprises bonafide utterances and fake utterances from 13 speech generators, referred to as A07 to A19, which are different from fake generators in `Train' and `Development' subsets.
%
We follow the ASVspoof challenge guidelines and use the Equal Error Rate (EER) as the primary evaluation metric for the proposed models. Additionally, we report Accuracy (Acc.), F1 score, and AUC to facilitate performance comparison among the models.

\subsection{Experimental settings}
The proposed SDS system has been developed within the Pytorch framework. The architecture has been trained for 60 epochs in total, using Titan RTX 24GB GPU. The first 50 epochs have been used for the Stage 1 of the training. Then, the model is finetuned on two classes (fake/bonafide) for the remaining 10 epoches in Stage 2. The Adam method~\cite{Adam} is used for the optimization.


\subsection{Results and discussion}
To evaluate our proposed model, we first construct a baseline leveraging ResNet18 backbone for feature map extraction and an Entropy head for classification (This Entropy head presents the architecture mentioned in stage 2 of the training process in section~\ref{train}). The baseline is trained on two classes (fake and bonafide) using Cross-Entropy loss.
We also evaluated three other models, referred to as M1, M2, and M3.
M1 uses the architecture in the stage 2 which presents the depthwise-inception backbone as shown in right side of Fig.~\ref{F1} and the Entropy head. This model is train from scratch on two classes (fake and bonafide) using Cross-Entropy loss. 
We compare M1 with the baseline to evaluate the role of proposed DIN architecture.
M2 presents the architecture in the stage 1. Then, this model is fine-tuned in stage 2, but it is not applied in the stage 3.
We evaluate the role of contrastive training strategy with multiple classes and multiple loss functions in M2 model.
Finally, M3 is the proposed full model of Fig.~\ref{F1} including the entire three-stage training strategy which leads to the estimation of the bonafide distribution in stage 3.

Experimental results in Table~\ref{table:R1} indicate that M1 with DIN architecture outperforms the ResNet18 baseline over all metrics.
While M1 presents a complexity with 1.77 M parameters and 985 M FLOPS, ResNet baseline shows a much higher complexity with 11.18 M parameters and 1192 M FLOPS.
When the proposed contrastive training strategy in stages 1 and 2 is applied to M2, it significantly improves the EER performance by 2.6\%.
Applying stage 3 where the bonafide distribution is used to measure the Mahalanobis distances among utterances helps to further improve all metric scores.

\begin{figure}[t]
    \centering
    \includegraphics[width =1.0\linewidth]{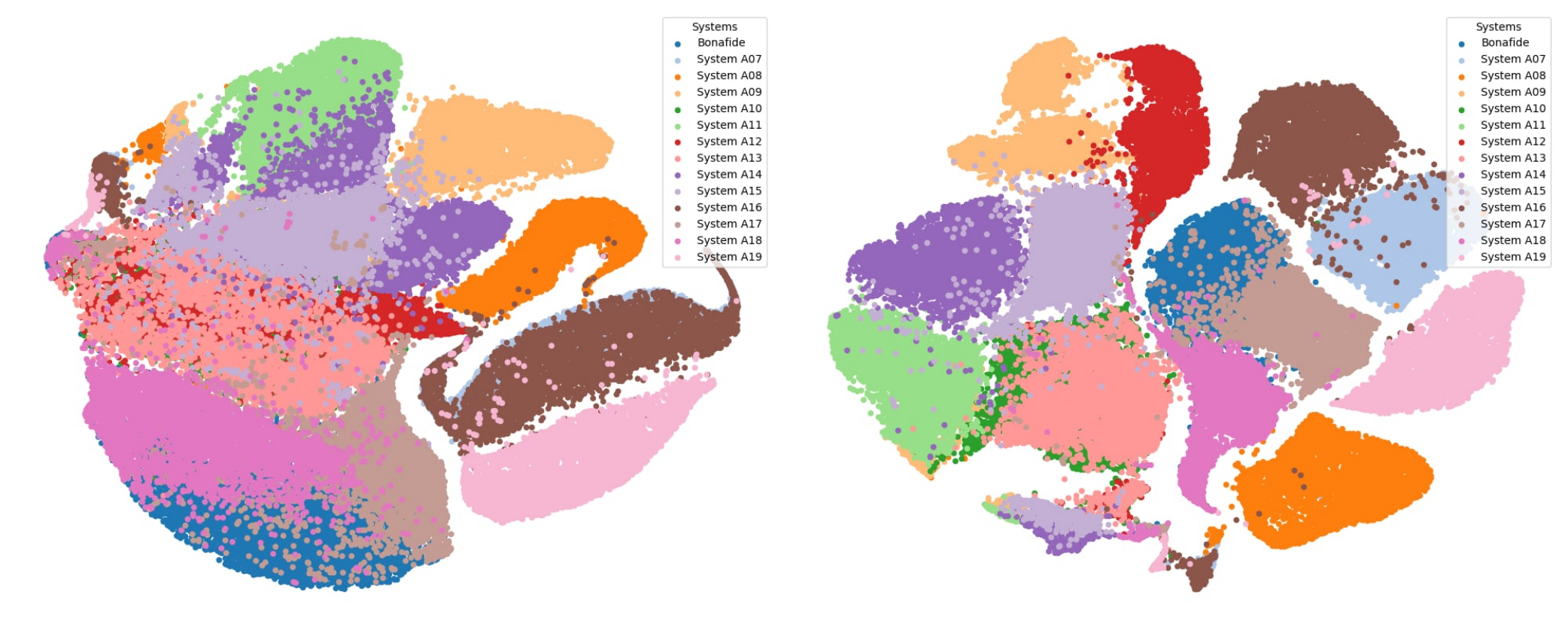}
           	\vspace{-0.2cm}
	\caption{TSNE-based visualization of the output feature maps extracted from the backbone (embedding $\boldsymbol{X}$ extracted from model M1 on the left side and extracted from model M3 on the right side) among testing bonafide and fake utterances (`Evaluation' subset with A07 to A19 generators)}
       	\vspace{-0.4cm}
    \label{F4}
\end{figure}

The visualization shown in Fig.~\ref{F3} indicates that the proposed three-stage contrastive training strategy is effective to separate the distributions of fake and bonafide utterances. This leads a smaller overlapping region between bonafide and fake utterances in M3 model compared with M1 model. 
The visualization shown in Fig.~\ref{F4} again proves that the contrastive training strategy is effective to separate bonafide utterances and fake utterances from different generators.

As a result, we achieve the single model M3 with the best EER score of 4.6\%, which outperforms submitted single systems in the challenge of ASV-2019 LA task (Top-4 single system in the challenges: T32 (4.92\%), T04 (5.74\%), T01 (6.01\%), and T58 (6.14\%)  in~\cite{top3}).
With a low complexity of 1.77 M parameters and Acc., F1, AUC scores of 95.4\%, 97.3\%, and 98.9\%, the model M3 shows strong potential for real-time deployment in DSD systems.
\section{Conclusion}
This paper has presented a deep-learning-based model for the task of deepfake speech detection. By combining the depthwise-inception network architecture and the three-stage contrastive training strategy, we achieve a low-complexity and high-performance single model (M3) which shows great for real-time DSD applications. 

\section*{ACKNOWLEDGMENTS}
EUCINF project is co-funded by European Union under grant agreement N°101121418.
Views and opinions expressed are however those of the author(s) only and do not necessarily reflect those of the European Union or the European Commission. Neither the European Union nor the granting authority can be held responsible for them.

Defame Fakes is funded by the Austrian security research programme KIRAS of the Federal Ministry of Finance (BMF).

\begin{figure}[h]
    \centering
    \includegraphics[width =1.0\linewidth]{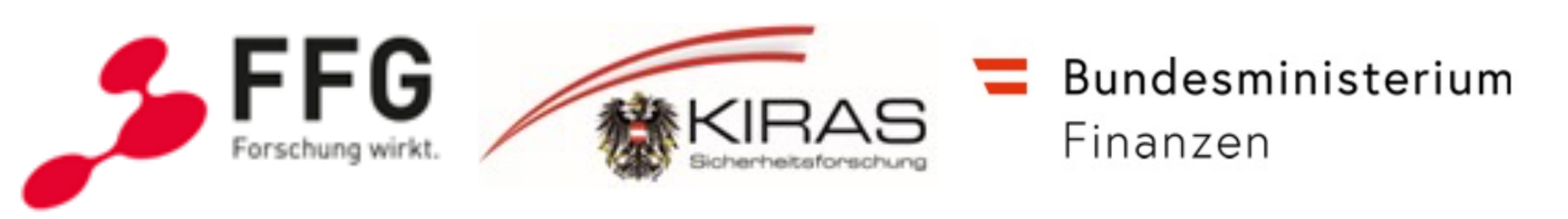}
           	\vspace{-0.2cm}
\end{figure}



\bibliographystyle{IEEEbib}
\bibliography{refs}
\end{document}